# genomepy: genes and genomes at your fingertips

Siebren Frölich[1,*], Maarten van der Sande[1], Tilman Schäfers[1], Simon J van Heeringen[1,*]

[1]Radboud University, Department of Molecular Developmental Biology, Faculty of Science, Radboud Institute for Molecular Life Sciences, 6525GA Nijmegen, The Netherlands
[*] Corresponding author

Contact information: siebrenf@science.ru.nl, s.vanheeringen@science.ru.nl

## Abstract

Analyzing a functional genomics experiment, such as ATAC-, ChIP- or RNA-sequencing, requires reference data including a genome assembly and gene annotation. These resources can generally be retrieved from different organizations and in different versions. Most bioinformatic workflows require the user to supply this genomic data manually, which can be a tedious and error-prone process.

Here we present genomepy, which can search, download, and preprocess the right genomic data for your analysis. Genomepy can search genomic data on NCBI, Ensembl, UCSC and GENCODE, and compare available gene annotations to enable an informed decision. The selected genome and gene annotation can be downloaded and preprocessed with sensible, yet controllable, defaults. Additional supporting data can be automatically generated or downloaded, such as aligner indexes, genome metadata and blacklists.

Genomepy is freely available at https://github.com/vanheeringen-lab/genomepy under the MIT license and can be installed through pip or bioconda.

# Introduction

Data analysis is increasingly important in biological research. Whether you are analyzing gene expression in a collection samples or transcription factor binding motifs in genomic atlases, you will need external information such as a reference genome or a specific gene annotation. For these types of data, there are three major providers: Ensembl[1], UCSC[2] and NCBI[3], and many model-system specific providers, such as GENCODE[4], ZFIN[5], FlyBase[6], WormBase[7], Xenbase[8] and more. Providers have different approaches to compiling genome assemblies and gene annotations, which affect formats, format compliance, naming, data quality, available versions and release cycle. These differences significantly impact compatibility with research[9], tools and (data based on) other genomic data.

You could try to find genomic data yourself, but there are many options. Ensembl, UCSC and NCBI each have FTP archives, web portals, and REST APIs, which you can use to search their individual databases. Alternatively, there are several tools that can be used to access some of these databases programmatically, such as ncbi-genome-download[10] and ucsc-genomes-downloader[11]. However, none of these can search, compare or download from all major genome providers data. Furthermore, downloading and processing genomic data manually can be tedious, error-prone, and poorly reproducible. Although the latter could be remedied by a data management tool, such as iGenomes[12], refGenie[13] or Go Get Data[14], most data managers still require the user to add new data manually.

We have developed genomepy to 1) find genomic data on major providers, 2) compare gene annotations, 3) select the genomic data best suited to your analysis and 4) provide a suite of functions to peruse and manipulate the data. Selected data can be downloaded from anywhere, and is processed automatically. To ensure reproducibility, data sources and processing steps are documented, and can be enhanced further by using a data manager. Genomic data can be loaded into genomepy, which uses and extends on packages including pyfaidx[15], pandas[16] and MyGene.info[17] to rapidly work with gene and genome sequences and metadata. Similarly, genomepy has been incorporated into other packages, such as pybedtools[18] and CellOracle[19]. Genomepy can be used on the command line and through the (fully documented) Python API, for a one-time analysis or integration in pipelines and workflow managers.

# Features of genomepy

The key features of genomepy are 1) providing an overview of available assemblies with the `search` function, 2) download and processing of a selected assembly, with the `install` function and 3) using assembly data through the Python API.

The `search` function queries the databases of NCBI, Ensembl, UCSC and GENCODE (caching the metadata), for text, taxonomy identifiers or assembly accession identifiers. The input type is automatically recognized and used to find assemblies that have the text in the genome names or various description fields, matches the taxonomy identifier or (partially) matches the assembly accession. The output of the `search` function is a table with rows of metadata for each assembly found. The metadata contains the assembly name and accession, taxonomy identifier, and indicates whether a gene annotation can be downloaded (or which of the four UCSC annotations) (see Figure 1A). Snippets of available gene annotation(s) can be inspected with the `annotation` function (Fig. 1B).

```
A $ genomepy search GRCh38
    name          provider accession        tax_id annotation species           other_info
                                             n r e k    <- UCSC options (see help)
    GRCh38        GENCODE  GCA_000001405.15  9606   ✓          Homo sapiens      GENCODE annotation + UCSC genome
    GRCh38.p13    Ensembl  GCA_000001405.28  9606   ✓          Homo sapiens      2014-01-Ensembl/2021-11
    hg38          UCSC     GCA_000001405.15  9606   ✓ ✓ x ✓    Homo sapiens      Dec. 2013 (GRCh38/hg38)
    GRCh38        NCBI     GCF_000001405.26  9606   ✓          Homo sapiens      Genome Reference Consortium
    GRCh38.p1     NCBI     GCF_000001405.27  9606   ✓          Homo sapiens      Genome Reference Consortium
    GRCh38.p2     NCBI     GCF_000001405.28  9606   ✓          Homo sapiens      Genome Reference Consortium
    ...
    GRCh38.p13    NCBI     GCF_000001405.39  9606   ✓          Homo sapiens      Genome Reference Consortium
    GRCh38.p14    NCBI     GCF_000001405.40  9606   ✓          Homo sapiens      Genome Reference Consortium
    ^
    Use name for genomepy install

B $ genomepy annotation GRCh38.p13 | cut -f1 -d\;
  09:58:47 | INFO | Ensembl
  1         ensembl_havana  gene       1471765 1497848 .     +    .    gene_id "ENSG00000160072"
  1         ensembl_havana  transcript         1471765 1497848 .   +    .    gene_id "ENSG00000160072"
  09:58:51 | INFO | NCBI
  NC_000001.11   genomepy       transcript      11874    12227 .     +    .    gene_id "DDX11L1"
  NC_000001.11   genomepy       exon            11874    12227 .     +    .    gene_id "DDX11L1"

C $ genomepy install --annotation GRCh38.p13
  $ ls -hl --group-directories-first .local/share/genomes/GRCh38.p13/
  total 4.3G
  drwxr-xr-x 3 simon simon 4.0K Aug  5 09:42 index
  -rw-r--r-- 1 simon simon  68K Aug  5 09:36 assembly_report.txt
  -rw-r--r-- 1 simon simon  30M Aug  5 09:38 GRCh38.p13.annotation.bed
  -rw-r--r-- 1 simon simon 1.4G Aug  5 09:38 GRCh38.p13.annotation.gtf
  -rw-r--r-- 1 simon simon  25K Aug  5 09:40 GRCh38.p13.blacklist.bed
  -rw-r--r-- 1 simon simon 3.0G Aug  5 09:35 GRCh38.p13.fa
  -rw-r--r-- 1 simon simon 6.3K Aug  5 09:35 GRCh38.p13.fa.fai
  -rw-r--r-- 1 simon simon 3.1K Aug  5 09:35 GRCh38.p13.fa.sizes
  -rw-r--r-- 1 simon simon 2.0K Aug  5 09:36 GRCh38.p13.gaps.bed
  -rw-r--r-- 1 simon simon  564 Aug  5 09:38 README.txt
```

*Figure 1: Examples of the genomepy command-line tools. A) The `search` command searches providers for the given text, assembly accession or taxonomy identifier. B) The `annotation` command provides a sample of the available gene annotation files. C) The `install` command downloads the genome sequence, associated files and, optionally, gene annotation.*

An assembly name can be passed to the `install` function (Fig. 1C). The genome FASTA file is downloaded with the desired sequence masking level[20,21] and alternate sequences (softmasked and none by default, respectively). Alternate sequences reflect biological diversity and are often contained in reference assemblies. During sequence alignment however, similar reference sequences result in multiple alignment, leading to loss of data (as discussed in[22]). Additional filters may be passed to either include or exclude contigs (chromosomes, scaffolds, etc.) by name or regex pattern. Once processed, a genome index is generated using pyfaidx[15], as well as contig sizes and contig gap sizes.

Gene annotations come in a variety of recognized formats (GFF3, GTF, BED12). The `install` function will download the most descriptive format available, to output the commonly used GTF and BED12 formats. Contig names of the genome and gene annotation sometimes mismatch, which makes them incompatible with tools such as splice-aware aligners. Therefore, genomepy will attempt to match the contig names of the gene annotations to those used in the genome FASTA.

The `install` function can be extended with postprocessing steps via plugins. The options can be inspected and toggled with the `plugin` function. Briefly, the blacklist plugin downloads blacklists by the Kundaje lab[23] for the supported genomes. Other plugins support the generation of aligner indexes, including DNA aligner indexes for Bowtie2[24], BWA[25], GMAP[26] or Minimap2[27], and splice-aware aligners such as STAR[28] and HISAT2[29].

Assemblies not present on the major providers can be processed similarly by supplying the URLs or file paths to the `install` function. For data provenance and reproducibility, a

README file is generated during the installation process with time, source files, processing steps, and filtered contigs.

These features are available on both the command line interface and Python API. Additional features are available on the Python API, focused around two classes. The `Genome` class can be used to extract exact or random sequences from the FASTA, filter the FASTA and list the contigs, contig sizes and contig gaps. The `Annotation` class can be used to browse and filter the BED12 or GTF files as pandas dataframes[16], map gene identifiers to other types using mygene.info[17], map chromosome names to naming schemes of other major providers, and create a dictionary of any two GTF columns or attribute fields (to easily convert gene identifiers to gene names for instance).

# Conclusion

Obtaining suitable genomic data is a principal step in any genomics project. With genomepy, finding and downloading available assemblies becomes trivial. A genome, with the desired sequence masking, level of biological diversity, and contigs can be obtained with a single command. Gene annotations in GTF and BED12 format, and matching the genome, can similarly be obtained, with further options available in the Python API. Whatever install options you choose are logged for reproducibility, allowing you to start your analysis with confidence.

# Availability and implementation

The functionalities of genomepy are available from a command line interface, aimed at ease of use and integration in automated pipelines. Extended functionality is accessible using a Python application programming interface. The tool is freely available under the MIT license and can be installed using Bioconda[30], pip[31], or directly used in workflows with our Docker[32] image or snakemake[33] wrapper. Code and documentation are available on GitHub (https://github.com/vanheeringen-lab/genomepy) and GitHub-pages (https://vanheeringen-lab.github.io/genomepy/), respectively.

# Acknowledgements

The authors would like to thank Vicky Luna Velez for designing the genomepy logo, Dohoon Lee and Jie Zhu for contributing to the genomepy code, as well as the many GitHub users who provided feedback, ideas and suggestions.

# Funding

Netherlands Organization for Scientific Research [NWO grant 016.Vidi.189.081 to S.J.v.H.].